# Understanding High-Temperature Chemical Reactions on Metal Surfaces

*Pai Li, Xiongzhi Zeng, and Zhenyu Li\**

Hefei National Laboratory for Physical Sciences at the Microscale, University of Science and Technology of China, Hefei, Anhui 230026, China

**Abstract**

Chemical reactions on metal surfaces are important in various processes such as heterogeneous catalysis and nanostructure growth. At moderate or lower temperatures, these reactions generally follow the minimum energy path and temperature effects can be reasonably described by a harmonic oscillator model. At a high temperature approaching the melting point of the substrate, general behaviors of surface reactions remain elusive. In this study, by taking hydrocarbon species adsorbed on Cu(111) as a model system and performing extensive molecular dynamics simulations powered by machine learning potentials, we identify several important high-temperature effects, including local chemical environment, surface atom mobility, and substrate thermal expansion. They affect different aspects of a high-temperature surface reaction in different ways. These results deepen our understanding of high-temperature reactions.

**Introduction**

To study the mechanism of a complex surface process, such as graphene growth,[1,2] it is essential to understand the involved elementary reactions. Temperature plays an important role in these chemical reactions. At a low-to-moderate temperature, a reaction roughly follows the minimum energy path (MEP) with almost unperturbed crystalline surface structures. The corresponding thermal fluctuations are small and can thus be reasonably described with a harmonic approximation. Accordingly, the reaction rate constant can be estimated from the harmonic transition state theory.[3] When the temperature becomes high enough, we may have a melting instead of a crystalline surface. There is not yet a general picture available for such high-temperature surface reactions. It is desirable to systematically study high-temperature effects on chemical reactions since many of them take place at high temperatures.[4–6]

To characterize a chemical reaction, we are interested to know the concentrations of reactants, diffusivities of reactants and products, and the reaction rate constant, which all depend on details of atomic-scale interactions. Notice that, for reactions on surfaces, concentrations refer to coverages of different adsorbates, which can be defined in a unified way for both crystalline and melting surfaces using the density of adsorption sites in the crystalline-surface model as a reference. At a high temperature, it is usually not practical to reveal atomic details of a surface reaction experimentally. Theoretically, high-temperature reactions can be studied by directly sampling all relevant configurations. With enhanced sampling techniques, excess chemical potential can be obtained[7] to estimate concentrations of different adsorbate species. Mean square displacement (MSD) predicted by molecular dynamics (MD) simulations can be used to obtain the diffusivity of a specific species on the surface.[8] Reaction rate constants can be estimated after the potential of mean force is obtained along the reaction coordinate.[9] More importantly, by comparing results from these sampling models and those from the harmonic crystalline-surface model, high-temperature effects can be studied to deepen our understanding of high-temperature reactions.[10]

The sampling of vast configuration space is computationally very expensive. In principle, simulations can be sped up by using empirical force fields to substitute computationally more expensive density functional theory (DFT) or other quantum chemical methods in energy and force evaluation. However, it is usually very challenging to obtain an accurate force field for chemical reactions on a substrate. One possible solution is constructing a suitable machine learning model to calculate energy and force. Machine learning techniques, such as neural network potential[11,12] and Gaussian approximation potential (GAP),[13] have been successfully applied in many fields of chemical science.[14–16] In principle, it is always desirable to obtain an all-in-one machine-learning potential that can be used in all chemical environments. However, one should notice that improving transferability can also be computationally expensive. If only a limited number of specific systems are interested, instead of building an all-in-one potential, an alternative protocol is to train a specific potential for each system.

In this study, some reactions of hydrocarbon on the copper surface are investigated. We have studied high-temperature effects on reaction rate constant previously.[10] Here, we focus more on the surface concentration and diffusivity of some reactant and product species ($H/C/CH/CH_3/C_2$). All these species are important in graphene growth.[17–19] Machine learning driven MD simulations are performed at different temperatures with a GAP trained for each system. By comparing results from the sampling model and those from the crystalline-surface model, we identify several important high-temperature effects. At a high temperature, the local chemical environment of an adsorbate can be significantly different from that on a crystalline surface. Situations leading to a similar local chemical environment at low and high temperatures are discussed. While the reaction rate constant is mainly determined by the local chemical environment, diffusion of an adsorbate at a high temperature can also be strongly affected by the mobility of surface atoms. Thermal expansion of substrate materials at a high temperature can significantly change adsorbate-surface interaction and thus the equilibrium adsorbate concentration. These results provide us with useful insights in understanding the general behavior of adsorbates and their reactions on surfaces.

**Computational Details**

Electronic structure calculations were performed with DFT implemented in Vienna *ab initio* simulation package (VASP),[20,21] using the general-gradient-approximation (GGA) exchange-correlation functional parametrized by Perdew, Burke, and Ernzerhof (PBE).[22] The DFT-D2 model[23] was used to describe van der Waals (vdW) interactions. The projector augmented wave method[24] was used to describe core-valence interaction. A kinetic energy cutoff of 400 eV was used for the plane-wave basis set. The climbing image nudged-elastic-band (CI-NEB) method[25] was used for MEP identification and transition state location. Except as otherwise specified, a 4×4 Cu(111) surface slab model with five Cu layers and a vacuum layer thicker than 15 Å was adopted. The bottom Cu layer was fixed in geometry optimization and MD simulations.

For each GAP potential, about 2150 structures were used as the training set, which was mainly generated in an iterative way. First, about 50 structures were obtained from a high-temperature (3000 K) *ab initio* MD simulation to generate an initial version of GAP potential. With this GAP, relatively long MD simulations were performed at a lower temperature. 50 structures were then extracted from this MD trajectory to extend the training set with DFT calculations. Better GAP can be obtained to perform new MD simulations and generate new training-set structures. In each iteration, the temperature of MD simulations was changed and more structures were collected at the temperature close to the experimental temperature of graphene chemical vapor deposition (CVD) growth (Figure S1). Principal component analysis suggested a much better representation of the configuration space was obtained by our iterative approach to generate the training set compared to directly using high-temperature *ab initio* MD. To further improve the quality of each GAP, about 100 structures generated by randomly inserting the adsorbate into the surface slab system were also added to the training set. More details about GAP training can be found in the Supporting Information. Both energy and force calculations for test structures (Figure S2) and phonon band structure calculations (Figure S3) indicated that GAP potentials generated in such a way are very accurate.

The atomic simulation environment (ASE)[26] as a Python library was adopted to perform the GAP MD. Widom insertion and Kirkwood coupling parameter simulations were performed with ASE and homemade codes. For each species, we performed a 2-ns GAP-MD simulation at six temperatures ranging from 600 to 1600 K. For C and $C_2$, 10-ns trajectories were obtained for better statistical convergence, since they have low diffusivities at relatively low temperatures. NVT ensemble and Anderson thermostat were adopted. Timesteps of 0.5 and 2 fs were used for systems with and without H, respectively.

To characterize configurations in MD trajectories, coordination number (CN) and weighted height (WH) of the adsorbate were defined. CN records the number of Cu atoms surrounding an atom in the adsorbate, which is defined as[27]

$$CN(i) = \sum_j \frac{1-\left(\frac{r_{ij}}{r_0}\right)^n}{1-\left(\frac{r_{ij}}{r_0}\right)^m} \quad (1)$$

where *i* denotes the chosen center atom in the adsorbate, *j* is the index of Cu atoms within 5 Å from the adsorbate, and $r_0$ is the switching parameter which was set to be 2.4 Å for Cu-C and 2.0

Å for Cu-H. The exponential factors $n$ and $m$ were set to 18 and 36, respectively. WH records the height of an adsorbate atom relative to the surrounding Cu atoms

$$WH(i) = \sum_j \left[ \frac{1 - \left(\frac{r_{ij}}{r_0}\right)^n}{1 - \left(\frac{r_{ij}}{r_0}\right)^m} \times (\vec{r}_{ji} \cdot \hat{z}) \right] / CN(i) \qquad (2)$$

where $\hat{z}$ is the unit vector along the $z$ direction.

**Results and Discussion**

To characterize a chemical reaction on a surface, the reaction rate constant should be determined. At a relatively low temperature, the chemical reaction follows the MEP and the reaction rate constant can be estimated via the harmonic transition state theory. At a high temperature, paths significantly different from the MEP may become important. An explicit sampling of structures in these paths is required to estimate the reaction rate constant, which can be realized via an MD simulation if a reaction coordinate can be identified. As shown in our previous *ab initio* MD studies,[10] if there is a strong steric hindrance effect, a high-temperature reaction will effectively follow the MEP and the rate constant can still be estimated from the harmonic transition state theory based on a crystalline-surface model.

In addition to the reaction rate constant, the reaction rate also depends on the concentration of reactants. At the same time, the role played by a specific reaction in a complex chemical process also depends on the diffusivity of both reactants and products. Therefore, the concentration and diffusivity of adsorbates on the surface are also very important to fully characterize surface reactions. However, estimating diffusivity and equilibrium concentration from MD simulation can be more challenging compared to estimating reaction rate constant. Statistical convergence is usually difficult to reach, especially in the estimation of adsorbate diffusivity. Therefore, we use GAP models to speed up MD simulations in this study.

**Equilibrium concentration estimation.** If a thermodynamic equilibrium is reached, the concentration of different adsorbates on the surface is determined by the adsorption free energy ($G_{ad}$) or more precisely its derivative to the number of adsorbates, i.e. the chemical potential ($\mu$). Clearly, when the concentration is high, $G_{ad}$ has a contribution from adsorbates' lateral interaction, which can be described using techniques such as cluster expansion.[28,29] In this study, we focus on the diluted limit with negligible adsorbate-adsorbate interaction,[18,30] which is the case in graphene

CVD growth on Cu surface.[17] Notice that, driven by configuration entropy, adsorbates prefer to be homogeneously distributed on the surface with large adsorbate-adsorbate distances.

The chemical potential of adsorbates on the surface can be divided into two parts: the ideal-gas part and the excess chemical potential ($\mu^{ex}$) describing the interaction between adsorbates and the substrate. The excess chemical potential can be obtained using the Widom insertion method[31]

$$\mu^{ex} = -k_B T \times ln\langle \exp(-\phi/k_B T)\rangle \tag{3}$$

where $\phi$ is the interaction between the inserted trial particle and the environment (which is the Cu substrate in this case), the angle bracket denotes an average on test insertions. To improve sampling efficiency, test insertions only target a subspace of the whole simulation box, which includes the first two layers of Cu atoms and a vacuum layer above. Since adsorbates strongly prefer to be adsorbed on the surface, such a restriction on the insertion space does not change the final result of the total chemical potential of the corresponding adsorbate. Test calculations with the Kirkwood coupling parameter method[9] are performed as a comparison to make sure that our results are well converged.

The ideal-gas part of the chemical potential can be written as the summation of $k_B T \times \ln c$ and the ideal-gas chemical potential of a reference state ($\mu_0^{id}$) with a concentration $c_0$. Notice that $c$ and $c_0$ are concentrations in the three-dimensional insertion space, which can be uniquely mapped to adsorbate concentrations or coverages $\lambda$ and $\lambda_0$ if the simulation box size is specified. We choose $c_0$ as the number of adsorption sites in the crystal-surface model divided by the volume of the insertion space, which makes $\lambda_0=1$.

With excess chemical potential and ideal-gas chemical potential at the reference state obtained, the adsorbate concentration at a specific chemical potential $\mu$ (determined by the experimental chemical environment) can be written as

$$\lambda = \exp\left(-\frac{\mu_0^{id} + \mu^{ex} - \mu}{k_B T}\right) \tag{4}$$

The stronger the adsorbate-substrate interaction, the lower the excess chemical potential and the higher the adsorbate concentration. Notice that, at a non-equilibrium steady state, the chemical potential $\mu$ of one species is not a constant. In such a case, an effective chemical potential can be defined[32] based on the kinetic network associated with the non-equilibrium steady state.

The estimated concentrations of $C/C_2/H/CH/CH_3$ from MD configuration sampling at different temperatures are shown in Figure 1. There is a fast increase of the equilibrium concentration with temperature for C/$C_2$/CH/$CH_3$. When the temperature is relatively low (600 or 800 K in this study), the equilibrium concentration is very low for all carbon-containing species, which gives a possible explanation for the experimental observation that high-quality monolayer graphene can hardly grow at low temperatures except that benzene or benzene-like molecules are used as the carbon precursor.[33] The concentration of H is less sensitive to temperature and it becomes slightly lower at higher temperatures because the dissociative adsorption of $H_2$ is weakly exothermic.[34] As a result, H coverage is much higher than that of carbon-containing species in graphene growth.

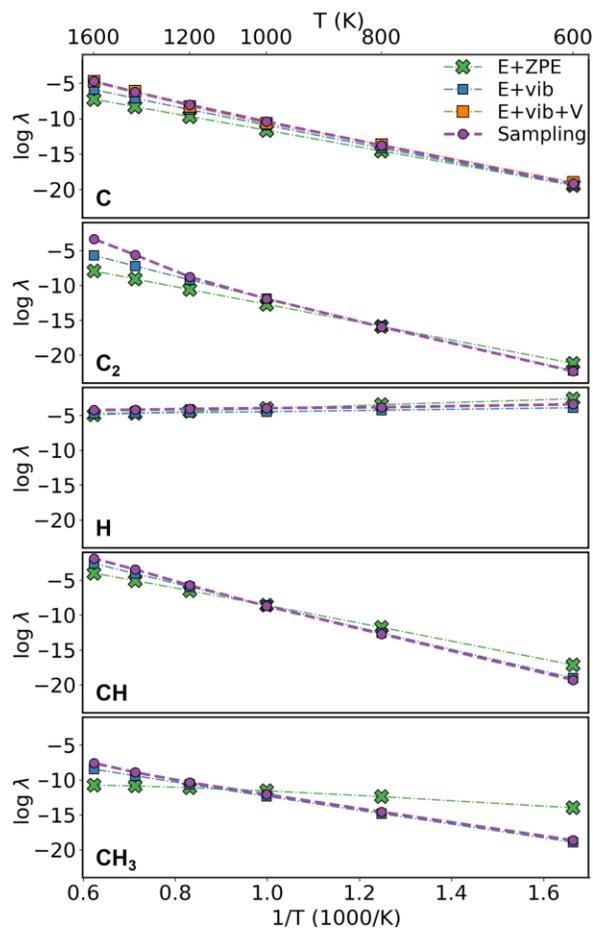

**Figure 1.** Adsorbate concentrations λ predicted by crystalline-surface and sampling models. "E+ZPE" means that adsorbate free energy is calculated based on adsorption energy with ZPE correction. "E+vib" means that the vibrational enthalpy and entropy contributions are also included. The result with a thermal volume expansion effect correction ("E+vib+V") is also provided for the C monomer. The chemical potential of each species is determined by assuming it is in equilibrium with graphene for C/$C_2$ and with a mixed gas phase ($P_{H_2}$=0.1 Torr, $P_{CH_4}$=10 Torr) for H/CH/$CH_3$.

At low temperatures, if the thermal fluctuation is harmonic, the adsorbate concentration can also be obtained based on the crystalline-surface model. $G_{ad}$ can be decomposed into three parts: the binding energy between an adsorbate and the substrate, vibrational contribution to free energies, and configurational entropy. The vibrational part can be further decomposed into three contributions: zero-point energy (ZPE), vibrational enthalpy, and vibrational entropy, in which the ZPE is independent of the temperature. At the dilute limit, the binding energy and vibrational contribution do not depend on the adsorbate concentration. In contrast, configurational entropy is solely determined by adsorbate concentration. Let $N$ and $n$ be the number of available adsorption sites and the number of adsorbates, then the configuration entropy is

$$S^{conf} = k_B \ln \frac{N!}{n!(N-n)!} = -[\lambda \ln \lambda + (1-\lambda)\ln(1-\lambda)]k_B N \tag{5}$$

where $\lambda = n/N$ is the adsorbate concentration. Stirling's approximation is applied here since $n$ and $N$ are large. At the diluted limit ($\lambda \ll 1$), $S^{conf}$ can be further simplified to

$$S^{conf} = n(1 - \ln \lambda)k_B \tag{6}$$

The chemical potential of the adsorbate is thus

$$\mu^{surf} = \frac{\partial G}{\partial n} = E_{all} - E_{sub} + \Delta G^{vib} + k_B T \times \ln \lambda \tag{7}$$

where $E_{all}$ and $E_{sub}$ are energies of the whole system and the substrate, respectively. $\Delta G^{vib}$ is the change of the vibrational contribution to the free energy upon adsorption. At a given chemical potential $\mu$, the adsorbate concentration is

$$\lambda = \exp\left(-\frac{E_{all} - E_{sub} + \Delta G^{vib} - \mu}{k_B T}\right) \tag{8}$$

As shown in Figure 1, at low temperatures, adsorbate concentrations predicted by the sampling model and by the crystalline-surface model described above agree well. This can also be considered as evidence that the statistics in configuration sampling are well converged. Notice that, in the literature, it is also a common practice to only include the ZPE part in the vibrational correction, which is exact only when the temperature is zero. Otherwise, the vibrational enthalpy and entropy contributions can be larger than the ZPE contribution, which can lead to a large difference between the sampling and crystalline-surface model predicted concentrations (up to three orders of magnitude). The deviation mainly occurs at the high-temperature side for C and $C_2$, at the low-temperature side for H, and at both sides for CH. The importance of vibrational free energy correction was also discussed for point defects in Si bulk system.[35,36]

**Diffusivity estimation.** A straightforward way to calculate the diffusion coefficient from an MD simulation is to calculate the MSD of diffusing particles, the slope of which is proportional to the diffusivity $D$ according to the Einstein relation[37]

$$D = \lim_{t \to \infty} \left[ \frac{1}{2dt} \langle [\vec{r}(t)]^2 \rangle \right] \tag{9}$$

where $d$ is the dimension in which diffusions take place (for surface diffusion, $d=2$), $t$ is time, and $\langle [\vec{r}(t)]^2 \rangle$ is MSD:

$$\langle [\vec{r}(t)]^2 \rangle = \frac{1}{N} \sum_{i=1}^{N} \langle [\vec{r}_i(t+t_0) - \vec{r}_i(t_0)]^2 \rangle \tag{10}$$

where $N$ is the number of diffusing particles, $\vec{r}_i$ is the position of the $i$-th particle, and $t_0$ is the reference time.

The simulation time should be long enough to reach a low statistic variance.[38] Notice that, although *ab initio* MD has been used to calculate Li-ion diffusivity in solid electrolytes,[39,40] it is computationally too expensive for $C_xH_y$ diffusion on Cu surface. First, on-surface diffusion takes place in 2 dimensions with fewer hopping sites. Second, there is typically only one diffusing particle in the simulation box with a low adsorbate concentration, which is not the case for Li-ions materials.[41] Therefore, the required simulation time can be as long as a few nanoseconds to reach a satisfying convergence for adsorbate diffusivity estimation, which is usually unaffordable in *ab initio* MD simulations.

In addition to the time-scale requirement, there is also a space-scale requirement on the size of the simulation box for surface diffusion simulations. For example, if a relatively small 4×4 supercell is used to represent the Cu surface, an unphysical atom-layer slide is frequently observed in MD simulations (Figure S6a). Actually, the slide of the whole top layer relative to the bottom layers in the 4×4 surface model has an energy barrier of 0.6 to 0.7 eV, which is comparable to species diffusion barriers. Such an unphysical surface layer slide phenomenon affects the diffusivity of adsorbates. To solve this problem, a large simulation box should be used (8×8 in this study). To perform long MD simulations with a large supercell, GAP is used to efficiently evaluate energy and forces. Diffusivities estimated from MSD are shown in Figure 2. As expected, diffusivity generally increases with temperature. Interestingly, there is a dip in the $C_2$ diffusivity-temperature curve, which will be explained later.

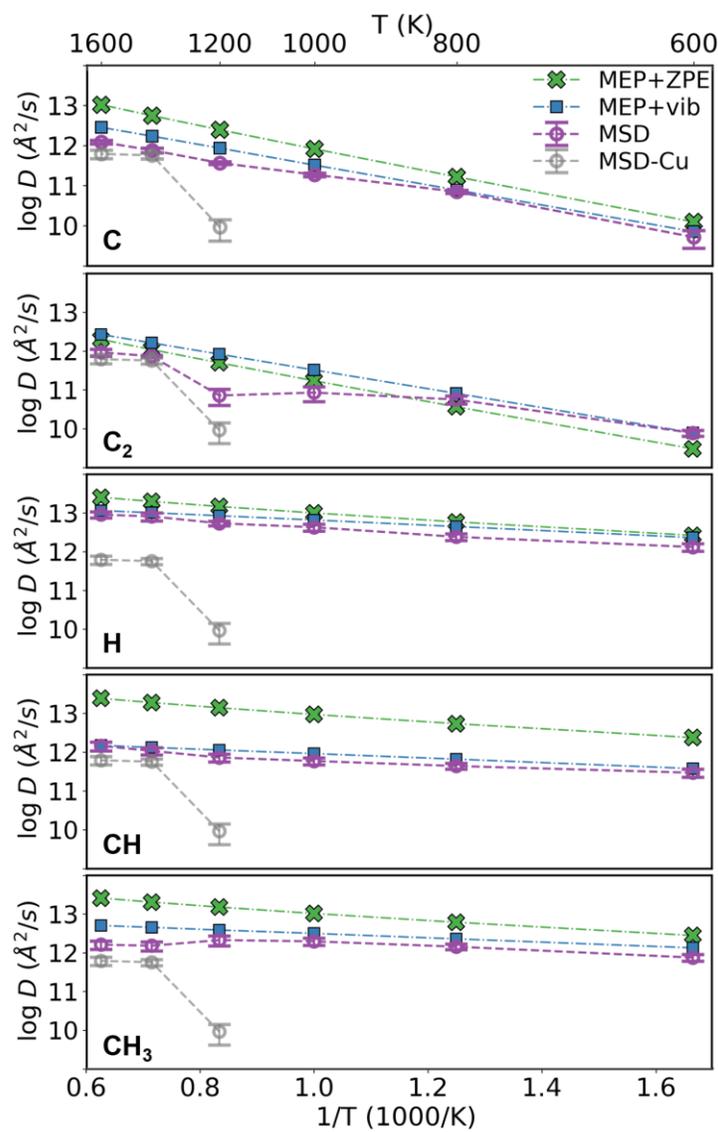

**Figure 2.** Diffusivity of different adsorbates (C/$C_2$/H/CH/$CH_3$) on Cu(111) surface estimated with different models. "MEP+ZPE" means that adsorbate hopping rate is calculated with the transition state theory, where energies along the MEP are corrected with the corresponding ZPE. "MEP+vib" means that free energy along the MEP is calculated with vibrational enthalpy and entropy contributions included. "MSD" means the diffusivity is obtained from MD simulations with MSD method. As a reference, the diffusivity of surface Cu atoms estimated from MSD is also shown at high temperatures.

At low temperatures, diffusion can be considered as hopping among different adsorption sites. Then, the rate can be more efficiently evaluated by calculating the hopping rate. According to the transition state theory, the tracer diffusion coefficient $D$ can be estimated by[37]

$$D = \frac{\Gamma a^2}{2d} \tag{11}$$

$$\Gamma = \frac{k_B T}{h} \times \exp\left(-\frac{\Delta G_{diff}}{k_B T}\right) \tag{12}$$

where $d$ denotes the dimension of the diffusion system (for surface diffusion, $d=2$), $a$ is the hopping distance, and $\Delta G_{diff}$ is the diffusion barrier along the MEP. The diffusion barrier can be calculated from the energy difference $\Delta E$ with a vibrational correction. Notice that, in frequency calculations, one typically should take adjacent substrate atoms into consideration. For example, if only the diffusing species is included, the correct vibrational mode with an imaginary frequency for the transition state of CH (Figure S5) cannot be obtained. In Table I, we list energy differences between the initial and the transition state ($\Delta E$) and ZPE, which are temperature independent. We can see that results from GAP and DFT calculations agree well, which again confirms the high quality of our GAPs. The full list of vibrational contributions can be found in Table S1.

**Table I.** Energy difference between the initial and the transition state ($\Delta E$) and ZPE for initial (IS), transition (TS), and final (FS) states for C/C$_2$/H/CH/CH$_3$ diffusion. Values before and after the slash are given by GAP and DFT, respectively.

|       | $\Delta E$ | ZPE(IS) | ZPE(TS) | ZPE(FS) |
|-------|------------|-------------|-------------|-------------|
| C     | 0.498/0.503 | 0.102/0.103 | 0.088/0.088 | 0.102/0.103 |
| C$_2$ | 0.489/0.521 | 0.212/0.213 | 0.190/0.187 | 0.212/0.213 |
| H     | 0.124/0.128 | 0.167/0.169 | 0.144/0.144 | 0.167/0.169 |
| CH    | 0.113/0.121 | 0.347/0.353 | 0.344/0.349 | 0.343/0.352 |
| CH$_3$| 0.087/0.093 | 0.873/0.905 | 0.888/0.907 | 0.891/0.904 |

As shown in Figure 2, at low temperature, the diffusivities predicted from MSD (the sampling model) and surface hopping (the crystalline-surface model) agree well. Similar to the diffusion case, the significance of different contributions of the vibrational free energy (ZPE, vibrational enthalpy, and entropy contributions) is compared. Only including the ZPE part of the vibrational contribution to free energy can lead to a significant error in diffusivity. For example, in almost the whole temperature range, the difference of CH diffusivity caused by vibrational enthalpy and entropy is as large as one order of magnitude (Figure 2). Further analysis indicates that such a difference is mainly from the vibrational entropy contribution.

**Effects of local chemical environment.** From above we can see that the sampling and crystalline-surface models always give similar results at low temperatures. However, at high temperatures, the difference can be either large or small. The most important effect that generates such a difference is the local chemical environment. In a previous study, we found that a reaction can effectively (locally) follow the MEP if the local chemical environment is similar to that of a crystalline surface.[10] For example, even at a high temperature with a melting surface structure, $CH_3$ adsorbed on Cu surface has a local chemical environment similar to $CH_3$ adsorbed at the bridge or hollow site on a crystalline Cu(111) surface. As a result, the reaction rate of $CH_3$ dehydrogenation at a high temperature can be estimated by the MEP-based crystalline-surface model.

Consistently, if we compare the equilibrium concentration of $C_2$ and $CH_3$, we find that the latter can be described by both sampling and crystalline-surface models. However, the former is underestimated by 2 or 3 orders of magnitude at high temperatures if the crystalline-surface model is used. This can be understood by checking the distribution of CN and WH for a C atom in $C_2$ in the sampling model (Figure 3). At low temperatures, CN and WH distribution are mainly around the minimum-energy structures, indicating sampled local chemical environments are similar to those in the crystalline-surface model. At high temperatures, as reflected by the lower WH, $C_2$ becomes closely wrapped by melting Cu atoms (Figure S8), which suggests a stronger adsorbate-surface interaction compared to that in the crystalline-surface model. As a result, a higher concentration compared to the crystalline-surface model is obtained. Notice that, even for $CH_3$, there is a small change of the local chemical environment at very high temperatures (1400 -1600 K) as indicated by the change of CN of C from 3 to 2. In the crystalline-surface model, this means a switch from hollow site adsorption to bridge site adsorption, which usually leads to a decrease in the adsorbate-surface interaction. However, due to the flexibility of Cu structures at high temperatures, the two Cu atoms connecting to C can devote a very large part of their bonding capability to C, which makes the substrate-adsorbate interaction can even slightly stronger compared to hollow site adsorption in the crystalline-surface model. Such a picture is confirmed by the slightly higher concentration predicted by the sampling model compared to those by the crystalline-surface model at high temperatures (Figure 1).

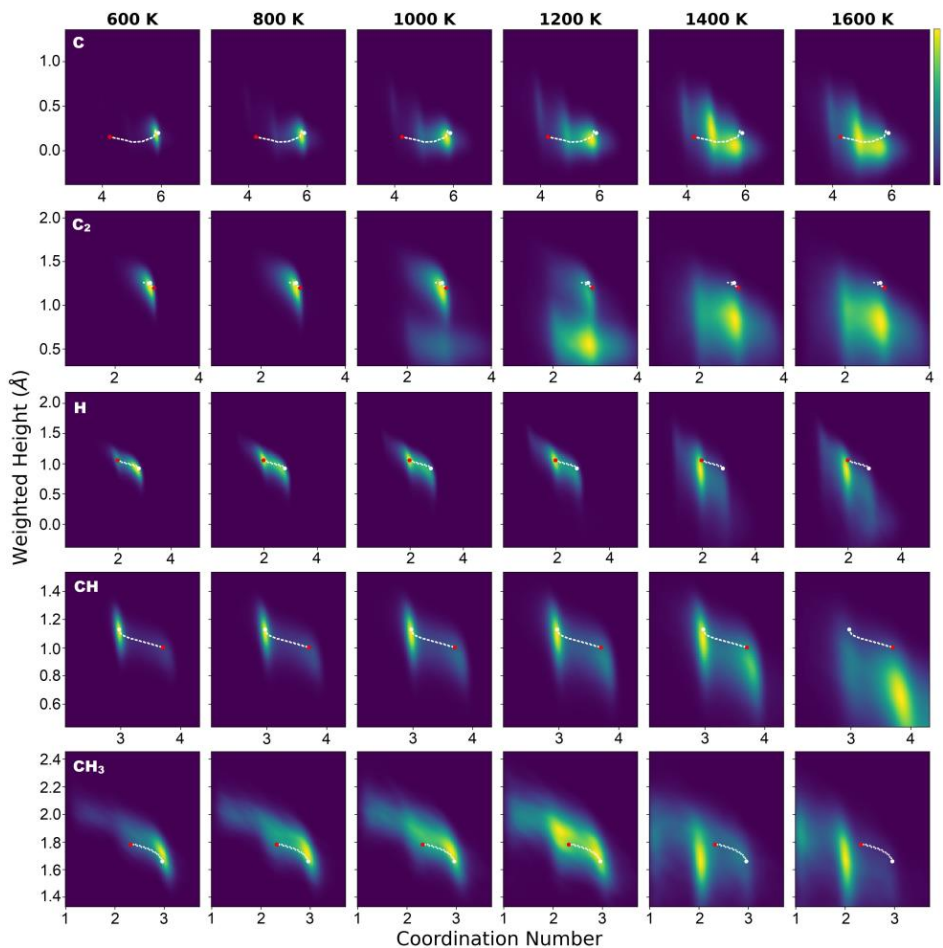

**Figure 3.** Distributions of CN and WH for C/C$_2$/H/CH/CH$_3$ adsorbed on Cu (111) surface in MD trajectories. For carbon-containing species, C is chosen as the center atom to calculate CN and WH. A brighter color means a larger possibility. White dots indicate IS/FS (two states overlap) and red dots indicate TS for adsorbate hopping on the crystalline surface. White dotted lines mark the hopping path.

Steric hindrance is the mechanism for CH$_3$ to largely keep the local chemical environment unchanged at high temperatures. As shown in Figure 3, the change of the local chemical environment is also small for C and H. However, their mechanisms are totally different since there is no steric hindrance effect in these two species. For C, CN in the crystalline-surface model is about 6, which corresponds to an octahedral site in the subsurface as also reported previously.[42] It means that C is already surrounded by Cu atoms, which is consistent with the fact that the WH value is close to zero. At high temperatures, the C atom remains to be surrounded by Cu atoms. Therefore, the change of the local chemical environment is small. For H, the key point is that the adsorbate atom is small. Therefore, even at high temperatures with a random Cu structure, H can only bond to 2 or 3 Cu atoms. Since the local chemical environment does not change much, we expect that the concentration of both C and H predicted from the sampling model will be similar to those predicted from the crystalline-surface model at high temperatures. This is the case for H and surprisingly it is not for C, which can be explained by the substrate thermal expansion effect and will be explained later.

The local chemical environment effect also has a strong influence on diffusivity. Here we focus on temperature T≤1200 K where the crystalline structure of Cu substrate is still well maintained (Figure 4a). Diffusion at even higher temperatures depends on the diffusion of substrate atoms and it is thus not a local effect. In the temperature range below 1200 K, as shown in Figure 3, the MEP hopping path well represents configurations sampled in MD simulations for C/H/CH/CH$_3$. As a result, diffusivity predicted from MD (the sampling model) agrees well with that from surface hopping (the crystalline-surface model) for these species. However, this is not the case for C$_2$, where configurations with WH much lower than the MEP structures become important at 1000 K and dominant at 1200 K. In such configurations, C$_2$ squeezes one Cu atom out and sinks into the top layer of the surface (Figure 4a). Such a sunk-adsorbate configuration is energetically 0.81 eV less favorable compared to the on-surface adsorption structure at zero temperature. Therefore, in the MEP of C$_2$ surface hopping, C$_2$ keeps on the surface. At 1200 K, both on-surface state and sunk-adsorbate state can be observed in the MD trajectory, while the latter dominates (Figure 4b). The sunk-adsorbate state predicted in MD has a much lower diffusivity since C$_2$ there is confined by Cu atoms in the first surface layer. This picture explains the sudden decrease of C$_2$ diffusivity predicted from the sampling model compared to the crystalline-surface model (Figure 2).

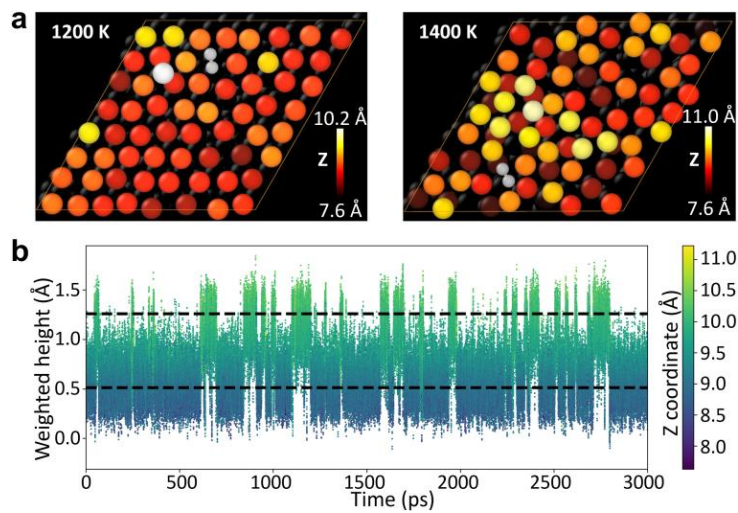

**Figure 4.** (a) Snapshots in MD trajectories for $C_2$ diffusion at 1200 and 1400 K. C atoms are shown in gray. Cu atoms are shown with a larger size, and higher atoms with brighter colors. For the 1200 K snapshot, we can clearly see one Cu atom is squeezed out from the surface layer. (b) Fluctuation of *WH* and the Z coordinate value of $C_2$ along the 1200 K trajectory. The upper and lower dashed lines correspond to the optimized on-surface and sunk-adsorbate states, respectively.

**Effects of substrate atom diffusion.** From the dynamic correlation function of adsorbate position (Figure S7), we can see that hopping is dominant in surface diffusion when T⩽1200 K. However, at 1400 or 1600 K, the diffusion behavior shows a smoothly-moving pattern. From a snapshot in the MD trajectory (Figure 4a), we can see that the surface is already melting at 1400 K. As a result, adsorption sites are not well defined and the hopping picture is not valid anymore. In fact, Cu atoms on the substrate are diffusive themselves. Therefore, when the surface is melting, diffusion can no longer be considered as a local property. It is not sufficient to just use the local chemical environment effect to explain adsorbate diffusion on a melting surface.

The diffusivity of Cu atoms themselves is close to $10^{12}$ Å$^2$/s at 1400/1600 K. Interestingly, the diffusivity of C/$C_2$/CH predicted from the hopping-based crystalline-surface model is close to the diffusivity of Cu. Therefore, these species can diffuse on the melting surface together with Cu atoms. Such a drifting effect can explain the abrupt increase of the MD predicted diffusivity of $C_2$ at 1400 K, where the sunk-adsorbate configuration does not exist anymore.

When the intrinsic adsorbate diffusivity (the value predicted by surface hopping) is notably higher than the diffusivity of Cu, the situation becomes more complicated. For bulky adsorbates with a strong interaction with the substrate, such as $CH_3$, friction from the substrate is expected, which decreases its diffusivity. As a result, a notable decrease of the MSD predicted $CH_3$ diffusivity is observed at 1400/1600 K. For small H adatom, even if the surface is melting, instantaneous diffusion channels may still be available which mitigate the friction from the surface. Therefore, the diffusivity of H is about $10^{13}$ Å$^2$/s at 1400/1600 K and it does not decrease toward the Cu diffusivity.

It is interesting to discuss the effect of substrate atom diffusion on equilibrium concentration and chemical reaction rate constant. Diffusion of Cu atoms is not expected to explicitly change the equilibrium concentration of adsorbates which is mainly determined by the adsorbate-substrate interaction and thus by the local chemical environment. However, the adsorbate-substrate interaction may also weakly depend on the diffusion of Cu, which can then implicitly change equilibrium concentration. For the chemical reaction rate constant, considering that both reactant/product and transition state will be affected by Cu diffusion in a more or less similar way, the effect of substrate atom diffusion on the chemical reaction rate constant is expected to be negligible.

**Effects of substrate thermal expansion.** At high temperatures, except that the melting surface can provide a different chemical environment and substrate atoms themselves are diffusive, there is another important effect for chemical reactions which is the thermal expansion of the whole substrate. Such an effect can be illustrated in the study of the equilibrium concentration of the C monomer. As already discussed, local chemical environment change can strongly change the equilibrium concentration, such as in the $C_2$ case. However, the difference of C concentration predicted by the sampling and crystalline-surface model cannot be described by a local chemical environment effect. Since C is wrapped by Cu atoms even at low temperatures and there is no significant chemical environment change at high temperatures.

In this case, the Cu volume expansion at high temperatures plays an important role. In the minimum energy structure of C adsorption on the Cu surface, we find that Cu atoms surrounding the octahedral adsorption site are repelled by C and the Cu crystalline structure is slightly distorted. This surface structure distortion can be considered as a local volume expansion. The C adsorption energy can then be decomposed into the energy penalty due to such a local volume expansion and the energy gain due to the interaction between C and Cu atoms. The competition between these two gives the optimal adsorption structure. To quantitatively illustrate such a picture, a simple model is built to describe the local volume expansion by artificially moving three surface atoms away from the center of the octahedral site (Figure 5a). Then, a local expansion ratio can be defined by comparing the volume of the octahedral interstitial space before and after the movement of these atoms. We calculate the energy penalty ($\Delta E_{expansion}$) and Cu-C interaction energy ($E_{C-Cu}$) as a function of the local expansion ratio. The summation of these two gives the minimum energy at the local expansion ratio of 1.23, which is close to the result from explicit geometry optimization (Figure 5b).

Since at high temperatures the substrate material will have a thermal expansion itself, the interaction strength between C and Cu in the crystalline-surface model is thus underestimated due to the overestimation of energy penalty. Such an effect can be taken into account by adding a compensation energy term $E_V$ in the crystalline-surface model to estimate the C coverage

$$\lambda = \exp\left(-\frac{E_{all} - E_{sub} + E_V + \Delta G^{vib} - \mu}{k_B T}\right) \tag{13}$$

$E_V$ is a function of temperature. To estimate $E_V$, we calculate the thermal expansion of bulk Cu via MD simulation in an NPT ensemble, where Langevin dynamics is used to control fluctuations in both the thermostat and the barostat. The obtained result is close to the quasi-harmonic approximation result at low temperatures (Figure 5c). Then, $E_V$ can be reasonably estimated as the energy penalty caused by the specific local space expansion equal to the bulk thermal expansion ratio. Such a protocol is rationalized by the fact that the change of Cu-C interaction energy on a surface with all Cu-Cu distances scaled by a global expansion ratio agrees well with what we obtain from the local expansion model (Figure S12). Using MD predicted bulk thermal expansion, we can estimate that $E_V$ is negligible (0.018 eV) at 500 K, but significant (0.369 eV) at 1600 K. When the $E_V$ correction is applied in the C concentration estimation (the E+vib+V model in Figure 1), the agreement between the sampling and crystalline-surface model becomes very well even at high temperatures.

Thermal expansion of substrate material is also expected to slightly affect the adsorbate diffusivity. At the same time, due to its similar effect on reactant/product and the transition state, the thermal expansion effect on reaction rate is expected to be negligible.

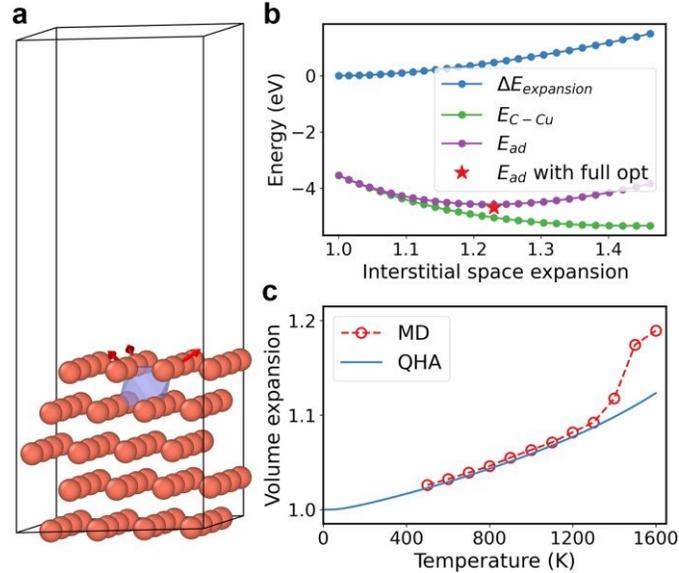

**Figure 5.** (a) Illustration of the local volume expansion due to C adsorption on Cu subsurface. (b) Energy penalty ($\Delta E_{expansion}$), Cu-C interaction ($E_{C\text{-}Cu}$), and their sum ($E_{ad}$) as a function of local volume expansion ratio. The red star gives the adsorption energy and local expansion ratio predicted by explicit geometry optimization. (c) The bulk volume expansion calculated via MD simulation and predicted with quasi-harmonic approximation (QHA) using Phonopy.[43]

**Experimental implications.** Results reported in this study provide insightful atomic details about graphene growth which cannot be obtained from an experimental means under the high-temperature growth conditions.[44] These information are useful for us in understanding the growth mechanisms. For example, the predicted low adsorbate concentration of carbon containing species on the surface can help us to understand why a very high temperature is required for graphene growth.

CVD growth of graphene on Cu surface is observed in experiments typically at a temperature close to the melting point of Cu. From chemical intuition, we expect that a high temperature is required to conquer the dehydrogenation barriers which are typically higher than 1 eV. However, it is not clear why the temperature should be as high as 1300 K. As shown in Figure 1, the equilibrium coverages of all carbon-containing species are positively related to temperature and they can be several orders of magnitude lower than those at the growth temperature even at 1000 K. Such low equilibrium coverage means graphene growth at a relatively lower temperature (for example, 1000 K) will be in one of the following two situations. If the pressure of the precursor gas is high, the low equilibrium coverages mean a very large carbon supersaturation and thus a poor sample quality. If a low pressure and a mild carbon supersaturation is maintained, then the growth rate will be too low. Therefore, a very high temperature is required to grow high-quality graphene with an acceptable growth rate.

The $C_2$ sunk-adsorbate state is reminiscent of the sunk of graphene islands on Cu surfaces.[45,46] It might also be connected with the formation of copper carbide.[47] Sinking of $C_2$ may be used as a nucleation mechanism to form a surface carbide layer at high temperatures. The thermal expansion effect revealed here is expected to be also important in bulk systems. For example, C can be dissolved in Ni with a high concentration. According to our test calculations, a C monomer at an interstitial site of bulk Ni metal also leads to a local expansion by a factor of 1.17, which causes an energy penalty of 0.40 eV. This is already similar to the C monomer in the Cu subsurface case. Therefore, if a crystalline Ni model is used to estimate C concentration at a relatively high temperature, compensation to the energy penalty should also be applied.

One thing we also want to emphasize here is that the protocol of this study is pretty universal, which can be used in the future to study more complicated systems. For example, in order to accelerate graphene growth, methods such as using Cu/Ni alloying substrate[48] and providing oxygen[49] or fluorine elements[50] are proposed. The role of these different elements in determining the equilibrium concentration of carbon-containing species should be carefully studied. The machine learning potential assisted sampling approach proposed here makes such studies possible.

**Conclusions**

In summary, equipped with well-trained GAPs, we have studied the diffusivities and concentration of five different species on the Cu surface. By comparing results from the sampling and crystalline-surface models, several important high-temperature effects are identified. Increasing temperature can change the local chemical environment of an adsorbate, generally from a support configuration to a wrap-around configuration. In three special cases, the local chemical environment does not change much. One possibility is that steric hindrance prevents the formation of a wrap-around configuration. Another possibility is that an embedded structure is already formed at low temperature, which is similar to the wrap-around structure at high temperature. The third possibility is that the adsorbate is very small, it can only bind to a very limited number of substrate atoms either at low or high temperatures. Local chemical environment change can significantly change the reaction rate constant, equilibrium concentration, and also diffusivity. At high temperatures, the substrate atoms become highly mobile, which has a significant effect on the high-temperature diffusivity of adsorbates. When the intrinsic diffusivity of the adsorbate is higher than the diffusivity of substrate atoms, there are two possibilities. For bulky adsorbate, an effective friction will be felt. However, for small adsorbates like H atom, instantaneous diffusion channels may still be available on a melting surface, which makes its diffusivity not be affected much by the surface atom mobility. Thermal expansion of substrate material can compensate the energy penalty for adsorption induced by a local volume expansion, which can change high-temperature adsorbate concentration. The results presented here provide a universal picture for high-temperature chemical reactions on metal surfaces.

ASSOCIATED CONTENT

**Supporting Information**.

The following files are available free of charge.

Details of vibrational corrections, thermodynamic sampling methods, and other supporting figures. The training sets and GAP potentials can be found at https://github.com/lipai/GAP-CxHy-on-Cu

AUTHOR INFORMATION

**Notes**

The authors declare no competing financial interests.

**Corresponding Author**

*Email: zyli@ustc.edu.cn

ACKNOWLEDGMENT

This work was partially supported by NSFC (21825302) and by USTC-SCC, SCCAS, Tianjin, and Shanghai Supercomputer Centers.

Supporting Information

# Understanding High-Temperature Chemical Reactions on Metal Surfaces

Pai Li, Xiongzhi Zeng, Zhenyu Li*

Hefei National Laboratory for Physical Sciences at the Microscale, University of Science and Technology of China, Hefei, Anhui 230026, China

(Email: zyli@ustc.edu.cn)

## 1. More computational details
### 1.1 Vibrational contributions on energy barriers

Three vibrational terms on energy barrier are zero-point-energy (ZPE), vibrational enthalpy, and vibrational entropy contributions:

$$\Delta_{ZPE} = E_{ZPE}^{TS} - E_{ZPE}^{IS} = \sum_{i \in TS} \frac{h v_i}{2} - \sum_{i \in IS} \frac{h v_i}{2}$$

$$\Delta_{vib\_enthalpy} = E_{vib}^{TS}(T) - E_{vib}^{IS}(T) = \sum_{i \in TS} \frac{h v_i}{\exp\left(\frac{h v_i}{k_B T}\right) - 1} - \sum_{i \in IS} \frac{h v_i}{\exp\left(\frac{h v_i}{k_B T}\right) - 1}$$

$$\Delta_{vib\_entrapy} = -TS_{vib}^{TS} + TS_{vib}^{IS}$$

$$TS_{vib} = k_B T \sum_i \left[\frac{h v_i}{k_B T \times \left[\exp\left(\frac{h v_i}{k_B T}\right) - 1\right]} - \ln\left(1 - \exp\left(\frac{-h v_i}{k_B T}\right)\right)\right]$$

where $v_i$ is the vibrational frequency of the *i*-th mode, $h$ is the Planck's constant, $k_B$ is the Boltzmann constant. TS and IS denote the transition and initial state, respectively.

### 1.2 Widom insertion to calculate the excess chemical potential

Since all hydrocarbon species considered in this study have a favored adsorption site on the top two Cu atom layers (surface hollow sites for H/CH$_3$, surface bridge sites for C$_2$, and subsurface interstitial site for C), we choose an insertion space containing surface/subsurface Cu atoms. In the vertical direction, we set the upper limit of the space (the upper limit of the lowest atom for multi-atom species) to be the average height of the top Cu atom layer plus one and half of the Cu-Cu layer spacing, the lower limit (the lower limit of the highest atom for multi-atom species) to be the average height of the second Cu layer minus half of the Cu-Cu layer spacing.

If we define the concentration of a species in this insertion space as the number ratio between species and Cu atoms in the thin volume, since the number of surface/subsurface Cu atoms equals the number of surface hollow sites, the 3D species concentration in the insertion space is identical to the 2D species coverage for H/CH$_3$. For other species, such a mapping can be obtained by simply applying a constant factor.

Given a specific species and temperature, we do the following steps:

*1.* Run GAP-MD for 110,000 steps at the given temperature with a pure-Cu surface model.

*2.* Abandon the first 10,000 steps and extract one structure from every 10 steps in the remaining 100,000 steps.

*3.* For each extracted substrate structure, randomly insert the given species into the insertion space, calculate and record the energy change $\phi$ due to insertion.

*4.* Repeat step 3 for 100 times.

Finally, we have 1,000,000 samples to calculate the average value of $\exp(-\phi/k_B T)$. The excess chemical potential can thus be acquired according to its definition

$$\mu^{ex} = -k_B T \times ln\langle\exp(-\phi/k_B T)\rangle$$

For CH/CH$_3$/C$_2$, we fixed the C-H/C-C bond length to the optimized value in the adsorption state. The energy difference caused by the bond-length constraints can be estimated via MD simulations of the system with and without bond length constraints at each temperature. It turns out to be small (about 0.14 eV for CH at 1000 K) and it is added as an energy correction to the excess chemical potential.

For monoatomic species (C and H), the insertion is performed by simply giving three evenly distributed random numbers to determine the insertion position. For diatomic species (CH and C$_2$), the position of one C atom is randomly given. After that, a random direction is given by providing three random numbers according to the normal distribution. The position of the other atom is determined by the bond length and the random direction. For CH$_3$, a random direction is given to determine the rotational symmetry axis and a random number is given to determine the rotation angle of H atoms around the symmetry axis.

The ideal-gas chemical potential at a reference density is defined as follows. For monomer, it contains only the translational part:

$$\mu_{trans}^{ideal} = k_B T \times \ln(\rho \times \Lambda_{sp})$$

where $\rho$ is the density of substrate atoms which slightly changed due to the volume expansion at a higher temperature, and $\Lambda_{sp}$ is the thermal de Broglie wavelength:

$$\Lambda_{sp} = (h^2/2\pi m k_B T)^{1/2}$$

For multi-atom species, it additionally contains the vibrational and rotational parts:

$$\mu_{vib}^{ideal} = \sum_i k_B T \times \ln\left(\frac{\exp(\frac{h v_i}{2 k_B T})}{1 - \exp(\frac{h v_i}{k_B T})}\right)$$

$$\mu_{rot}^{ideal} = \begin{cases} k_B T \times \ln\left(\frac{8\pi^2 I k_B T}{\sigma h^2}\right), & \text{if linear} \\ k_B T \times \ln\left[\frac{\sqrt{\pi I_A I_B I_C}}{\sigma}\left(\frac{8\pi^2 k_B T}{h^2}\right)^{3/2}\right], & \text{if nonlinear} \end{cases}$$

where $v_i$ is the $i$-th vibrational frequency of a multi-atom species, $I$ is the degenerate moment of inertia for a linear species, $I_{A/B/C}$ are the three principal moments of inertia for a non-linear species and $\sigma$ is the symmetry number.

**1.3 Kirkwood coupling parameter method**

For C and $C_2$, simulations based on the Kirkwood coupling parameter method were also performed. This method is based on a series of artificial systems, in which the interaction between the adsorbate and the Cu environment is scaled with a factor $\alpha$. We first calculate the excess chemical potential for the $\alpha=0.1$ system using Widom insertion:

$$\mu^{ex}(\alpha = 0.1) = -k_B T \times ln \langle \exp(\frac{-\phi \times 0.1}{k_B T}) \rangle$$

Then, the excess chemical potential for the real system $\alpha=1$ is calculated with thermodynamic integration:

$$\mu^{ex}(\alpha = 1) = \mu^{ex}(\alpha = 0.1) + \int_{0.1}^{1} \langle \phi \rangle_\alpha d\alpha$$

where $\langle \phi \rangle_\alpha$ is the ensemble average of $\phi$ for the artificial system with scaled factor $\alpha$.

To get $\langle \phi \rangle_\alpha$, we performed a series of MD simulations for different $\alpha$ ranging from 0.1 to 1 (the $\alpha=1$ system is the real system). To get the scaled interaction, we calculate forces for both the complete model ($F_{all}$) and the model with the adsorbate removed ($F_{subs}$). Then, the artificial forces are given by

$$F_{artif} = F_{subs} + \alpha \times (F_{all} - F_{subs})$$

For atoms in the adsorbate, $F_{subs}$ is zero. In this way, the interaction between Cu is kept but that between Cu and adsorbate is scaled with the factor $\alpha$. Meanwhile, the energy difference between two systems with and without the adsorbate ($\phi$) is recorded in each MD step to get its ensemble average $\langle \phi \rangle_\alpha$.

This artificial-force scheme for MD simulation can be directly used for the C monomer. For $C_2$, it scales also the C-C interaction which is incorrect. Here, we simply treat $C_2$ as a rigid body by fixing the C-C bond length in all MD simulations. The excess chemical potential obtained is finally corrected by applying the bond constrain correction at each temperature.

## 2. GAP training

For each hydrocarbon species, we first run *ab initio* MD at 3000 K for 50 steps. All these frames are used as the first batch of the training set, from which an initial GAP is obtained. Then, GAP-MD simulations are performed to obtain other configurations of the training set. To extract *N* structures at a temperature between 2000 and 600 K, we run a GAP-MD simulation at this temperature for 1000+100*$N$ steps. We uniformly extract one structure in every 100 frames for the last 100*$N$ steps. At temperatures below 600 K, the simulation steps are increased to 1000+1000*$N$ steps to extract a structure from every 1000 frames. DFT calculations for all these extracted structures are performed with VASP to get their energies and forces. The value *N* for each temperature is given in Figure S1a.

To train the less-accurate initial GAPs used for generating more structures in the training set, we use only SOAP descriptor[1] with an expansion of the neighbor density up to $l_{max}=6$, $n_{max}=10$, and cutoff=5 Å. For the final GAPs used for calculating diffusivity and coverage properties, we use distance_2b descriptor with cutoff=4 Å, angle_3b descriptor with cutoff=3 Å, and Soap descriptor with $l_{max}=8$, $n_{max}=12$, and cutoff=5.5 Å. The parameter atom_sigma is set to be 0.3 and 0.5 for systems with and without H involved, respectively.

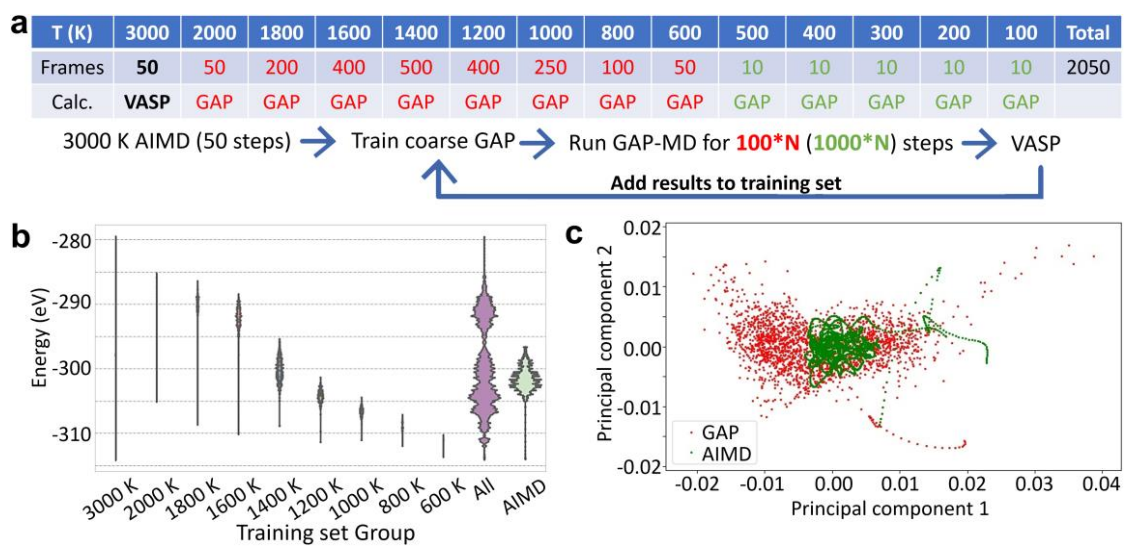

**Figure S1.** (a) The iterative training schemes. "*Frames*" in the table gives the number of structures extracted as a training set at each temperature. More structures were collected at temperatures close to the experimental temperature of graphene CVD growth. (b) Energy distribution in each batch of the training set, and in the whole training set (denoted as "*All*"). An ab initio MD trajectory at 1300 K denoted as "*AIMD*" with the same number of frames to the training set is presented as a comparison. (c) Principal component analysis on structures of the whole training set and that of the 1300 K AIMD trajectory. SOAP descriptor[1] is adopted to convert the structure into a vector for the principal component analysis.

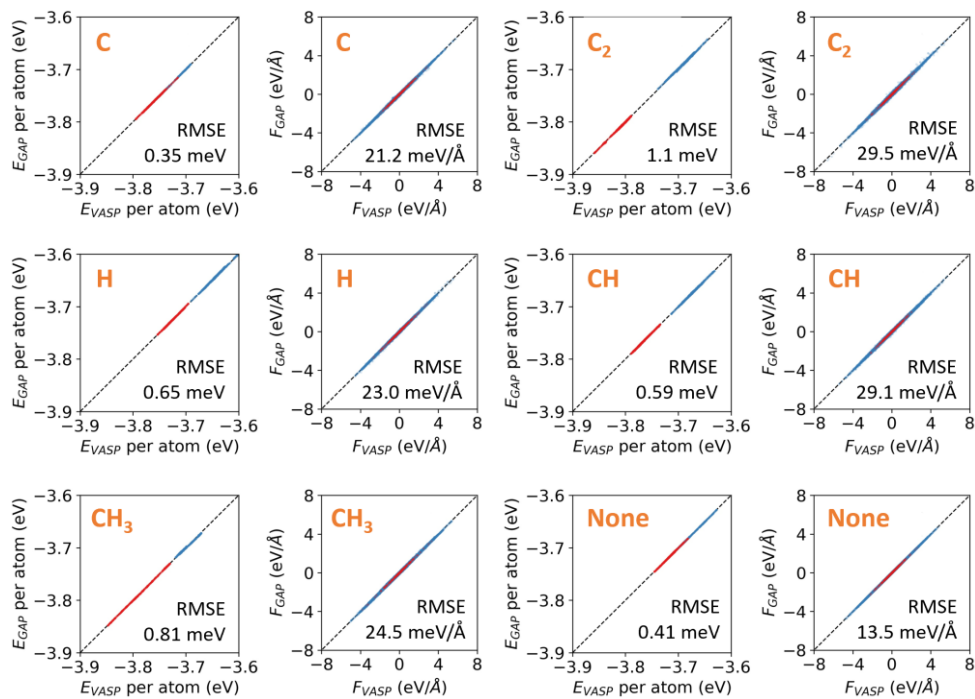

**Figure S2.** Energy and force error analysis for Cu substrate with adsorbate C/$C_2$/H/CH/$CH_3$ or without adsorbate (denoted as "None"). The test set obtained from an AIMD simulation at 1200 K is in red and that at 1400 K is in blue. All these results indicate high accuracy with very low root-mean-square errors (RMSE).

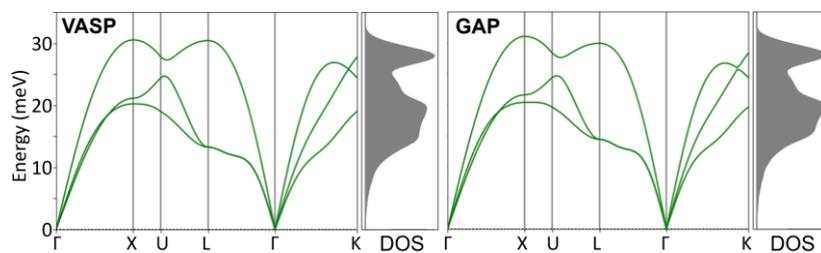

**Figure S3.** Cu phonon band structures calculated using VASP and GAP, respectively.

## 3. Crystalline surface model

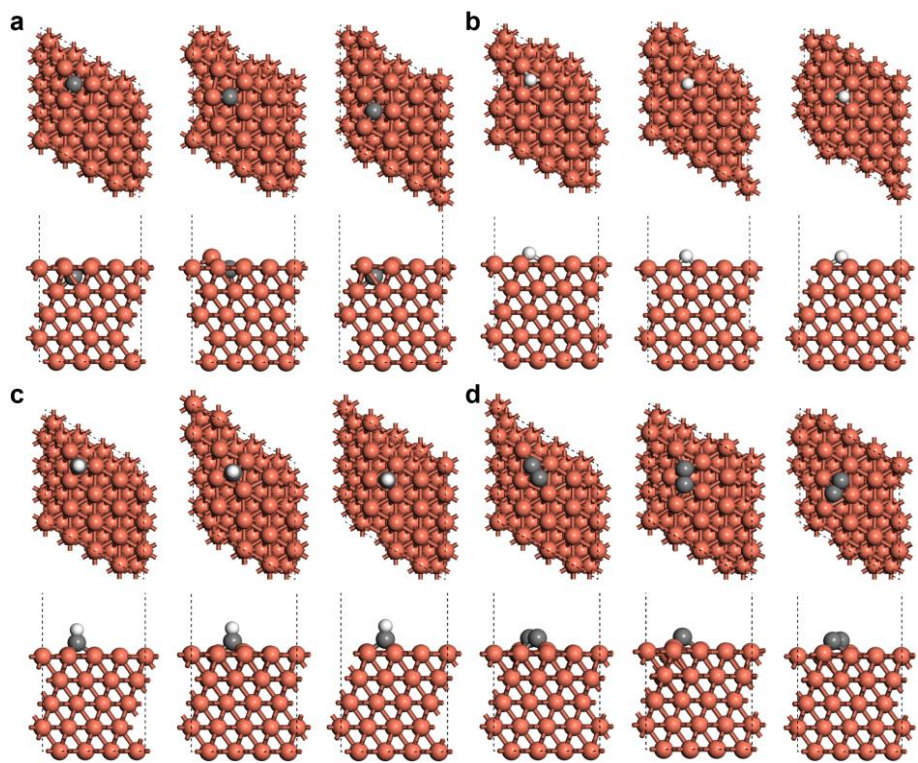

**Figure S4.** Initial/transition/final states in top and side views for (a) C, (b) H, (c) CH and (d) $C_2$ diffusion on Cu111 surface. Cu, C, and H atoms are shown in brown, grey, and white, respectively.

**Table S1.** Vibrational correction terms and the final free energy barriers for $C/C_2/H/CH/CH_3$ under different temperatures. T is the temperature in K. $\Delta$ZPE, $\Delta Cp$ and $\Delta(-TS)$ are the difference of ZPE, vibrational enthalpy, and vibrational entropy part between transition and initial states. $\Delta G_{vib}$ is the summation of these three terms. "*Barrier*" is calculated as the summation of the NEB barrier and $\Delta G_{vib}$. The substrate, except for the top layer, is fixed to calculate the vibrational frequencies.

|  | T | $\Delta$ZPE | $\Delta Cp$ | $\Delta(-TS)$ | $\Delta G_{vib}$ | Barrier |
|---|---|---|---|---|---|---|
|  | 600 | -0.023 | -0.033 | 0.062 | 0.006 | 0.504 |
|  | 800 | -0.023 | -0.049 | 0.101 | 0.029 | 0.527 |
| C | 1000 | -0.023 | -0.065 | 0.145 | 0.057 | 0.555 |
|  | 1200 | -0.023 | -0.083 | 0.192 | 0.086 | 0.584 |
|  | 1400 | -0.023 | -0.100 | 0.242 | 0.119 | 0.617 |
|  | 1600 | -0.023 | -0.116 | 0.295 | 0.156 | 0.654 |
|  | 600 | -0.033 | -0.027 | -0.022 | -0.082 | 0.407 |
|  | 800 | -0.033 | -0.043 | -0.012 | -0.088 | 0.401 |
| $C_2$ | 1000 | -0.033 | -0.059 | 0.004 | -0.088 | 0.401 |
|  | 1200 | -0.033 | -0.075 | 0.022 | -0.086 | 0.403 |
|  | 1400 | -0.033 | -0.092 | 0.044 | -0.081 | 0.408 |
|     | 1600 | -0.033 | -0.108 | 0.068 | -0.073 | 0.416 |
| H   | 600  | -0.024 | -0.024 | 0.031 | -0.017 | 0.113 |
|     | 800  | -0.024 | -0.042 | 0.061 | -0.005 | 0.125 |
|     | 1000 | -0.024 | -0.060 | 0.096 | 0.012  | 0.142 |
|     | 1200 | -0.024 | -0.077 | 0.134 | 0.033  | 0.163 |
|     | 1400 | -0.024 | -0.095 | 0.176 | 0.057  | 0.187 |
|     | 1600 | -0.024 | -0.111 | 0.220 | 0.085  | 0.215 |
| CH  | 600  | -0.002 | -0.049 | 0.144 | 0.093  | 0.206 |
|     | 800  | -0.002 | -0.067 | 0.212 | 0.143  | 0.256 |
|     | 1000 | -0.002 | -0.084 | 0.284 | 0.198  | 0.311 |
|     | 1200 | -0.002 | -0.101 | 0.360 | 0.257  | 0.370 |
|     | 1400 | -0.002 | -0.118 | 0.439 | 0.319  | 0.432 |
|     | 1600 | -0.002 | -0.136 | 0.520 | 0.382  | 0.495 |
| $CH_3$ | 600  | 0.004  | -0.042 | 0.08  | 0.042  | 0.129 |
|     | 800  | 0.004  | -0.061 | 0.129 | 0.072  | 0.159 |
|     | 1000 | 0.004  | -0.079 | 0.181 | 0.106  | 0.193 |
|     | 1200 | 0.004  | -0.096 | 0.237 | 0.145  | 0.232 |
|     | 1400 | 0.004  | -0.115 | 0.296 | 0.185  | 0.272 |
|     | 1600 | 0.004  | -0.133 | 0.358 | 0.229  | 0.316 |

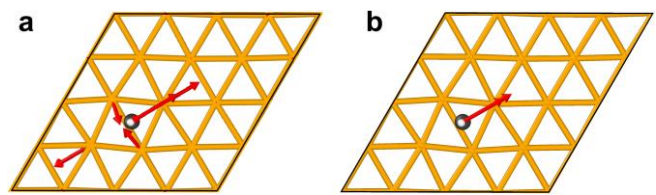

**Figure S5.** (a) Vibrational mode with an imaginary frequency in the transition state of CH diffusion with all surface atoms unfixed. (b) The corresponding model mode calculated with all Cu atoms fixed, where the vibrational frequency becomes real.

## 4. MSD and diffusion pattern

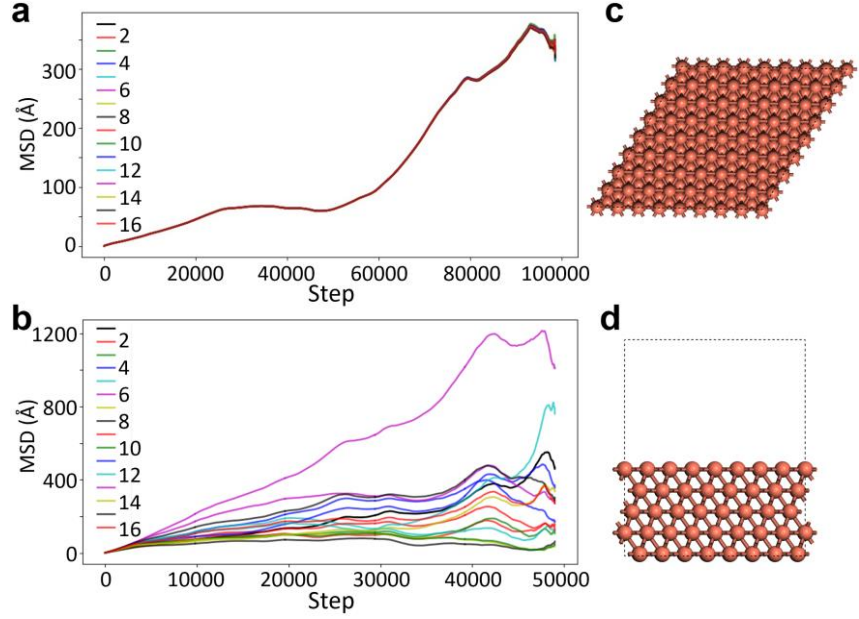

**Figure S6.** MSD of 16 Cu atoms in the top layer of a 4×4 surface obtained from test MD simulations at (a) 1000 and (b) 1400 K. The MSD of all atoms overlap at 1000 K because of the unphysical slide of the top layer due to the limited size of the surface model. At 1400 K, this phenomenon disappears since the surfaced is melted. (c) Top and (d) side view of the 8×8 surface model we used for the final MD calculations.

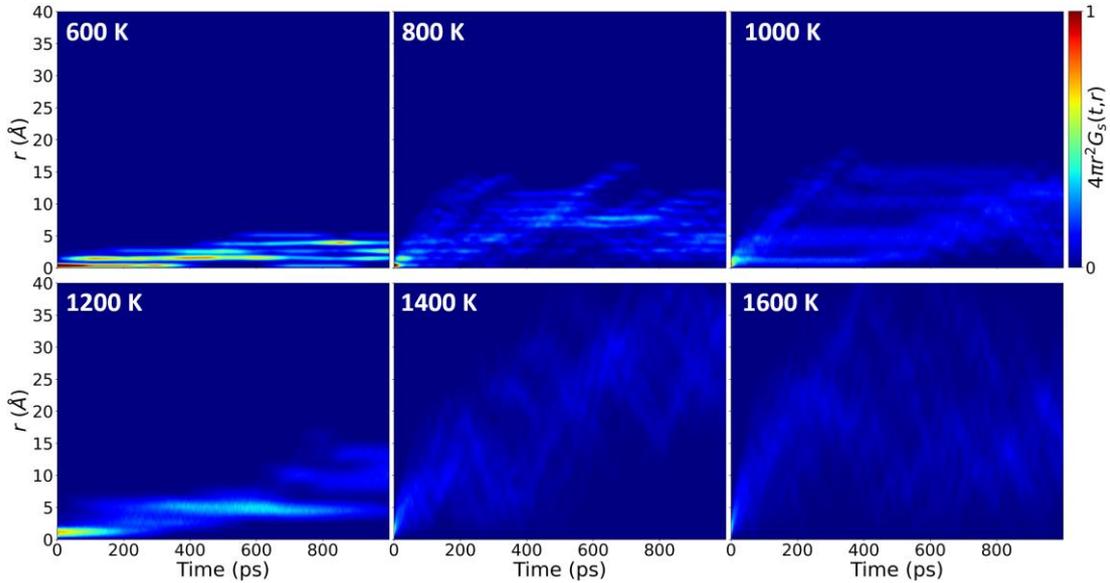

**Figure S7.** Self-part of the van Hove correlation function for $C_2$ diffusion at different temperatures. At a temperature lower than the Cu melting point, the correlation function shows a hopping picture. Otherwise, the Cu surface is melted and the diffusion behavior becomes Brownian motion on the liquid surface. The self-part of the Van Hove correlation function is calculated as follows:

$$G_s(r,t) = \frac{1}{4\pi r^2 N_d} \langle \delta(r - |r_i(t_0) - r_i(t+t_0)|) \rangle_{t_0}$$

where $\delta(\cdot)$ is the Dirac delta function, $N_d$ is the number of the diffusing species, $r_i(t)$ is the species position at time t. The angular brackets indicate ensemble average over the initial time $t_0$.

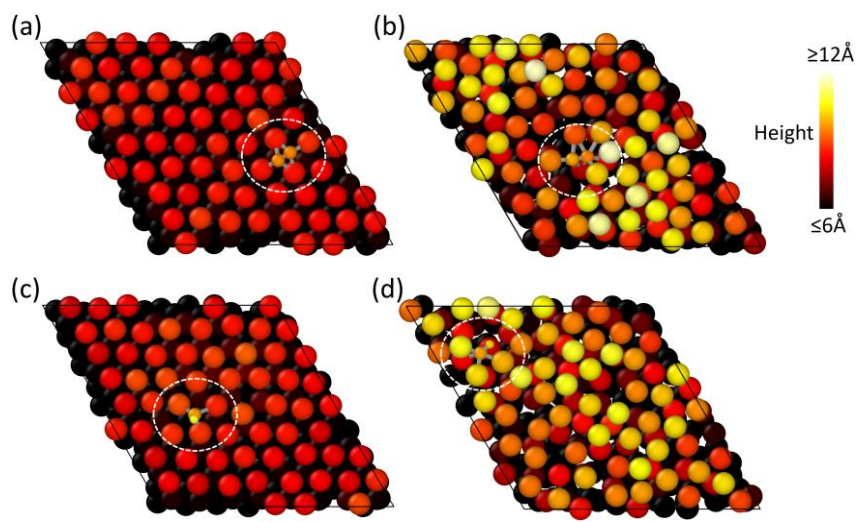

**Figure S8.** Snapshots of $C_2$ and CH in MD trajectories. (a) $C_2$ at 600 K; (b) $C_2$ at 1600 K; (c) CH at 600 K; (d) CH at 1600 K. Atoms' heights are distinguished with different colors. The largest balls are Cu atoms, the smallest balls are H atoms, and the other balls are C atoms.

## 6. Thermostat performance and convergence analysis

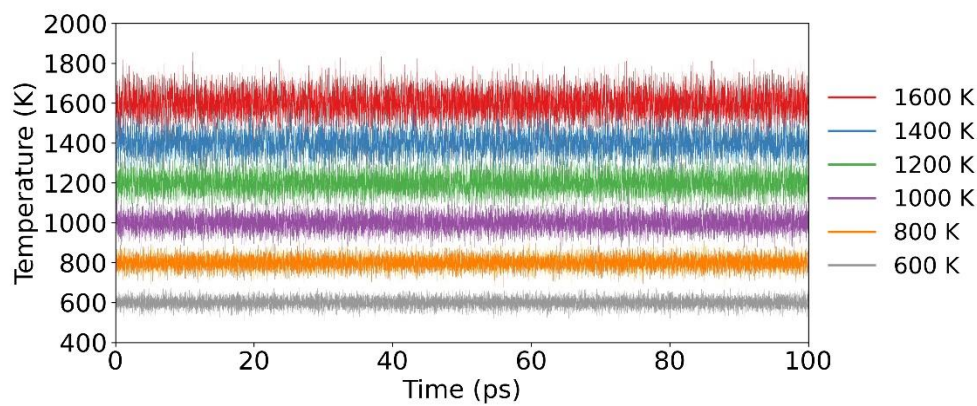

**Figure S9**. Temperature fluctuations in the last 100 ps for $C_2$ diffusion MD simulations at different temperatures.

To obtain converged diffusivity, it is suggested that the total mean square displacement (TMSD) in the simulation should be up to a few hundred Å$^2$.[2] Such a criterion is fulfilled in all our simulations for diffusivity (Figure S10), indicating the trajectories are long enough. For Widom insertion, the typical number of insertion attempts is $10^5 \sim 10^6$.[3,4] In this study, Widom insertion simulations are performed with $10^6$ attempts each. At the same time, we divide each simulation into five parts with equal steps and calculate the diffusivity or excess chemical potential based on each part. The obtained standard deviations are drawn as the error bars in Figure 2 and Figure S11, which are reasonably small.

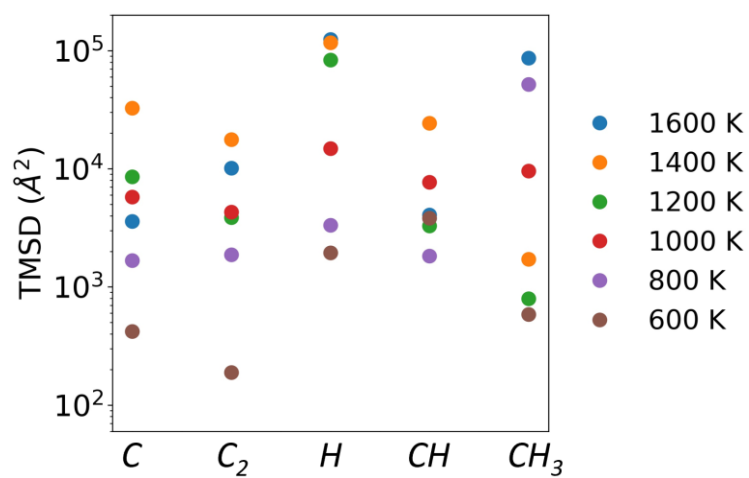

**Figure S10.** TMSD of adsorbates in MD simulations at different temperatures.

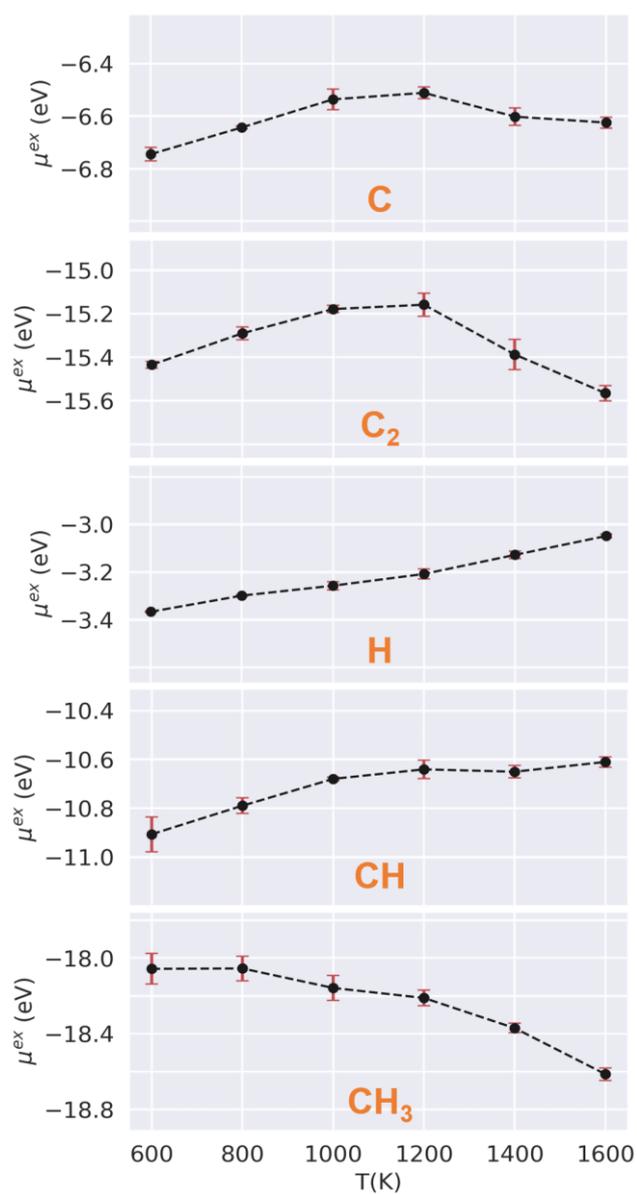

**Figure S11.** The excess chemical potential of $C/C_2/H/CH/CH_3$ adsorbates on Cu surface at different temperatures. The error bar gives the standard deviation of five independent sampling groups.

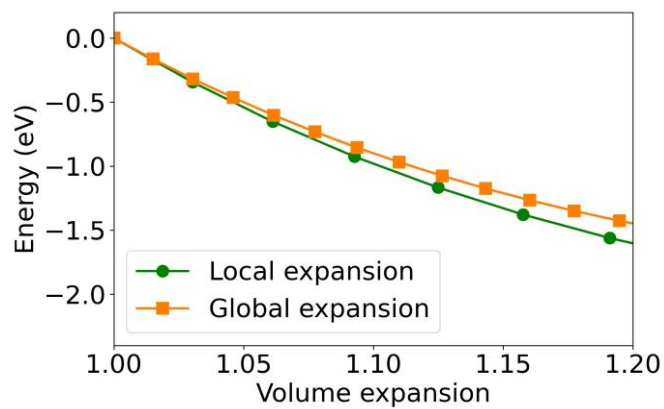

**Figure S12.** The adsorption energy of C monomer on expanded Cu substrate with local and global expansion models. In the global expansion model, the entire surface slab is expanded in the x/y/z directions. All Cu atoms are fixed and only the C monomer is optimized to get the interaction energy.